# Cellular Memristive-Output Reservoir (CMOR)

Wilkie Olin-Ammentorp

Colleges of Nanoscience & Engineering, SUNY Polytechnic Institute, Albany, NY, USA, WOlin-Ammentorp@sunypoly.edu

Karsten Beckmann

Colleges of Nanoscience & Engineering, SUNY Polytechnic Institute, Albany, NY, USA, kbeckmann@sunypoly.edu

Nathaniel Cady*

Colleges of Nanoscience & Engineering, SUNY Polytechnic Institute, Albany, NY, USA, ncady@sunypoly.edu

* Corresponding Author

## ABSTRACT

Reservoir computing is a subfield of machine learning in which a complex system, or 'reservoir,' uses complex internal dynamics to non-linearly project an input into a higher-dimensional space. A single trainable output layer then inspects this high-dimensional space for features relevant to perform the given task, such as a classification. Initially, reservoirs were often constructed from recurrent neural networks, but reservoirs constructed from many different elements have been demonstrated. Elementary cellular automata (CA) are one such system which have recently been demonstrated as a powerful and efficient basis which can be used to construct a reservoir.

To investigate the feasibility and performance of a monolithic reservoir computing circuit with a fully integrated, programmable read-out layer, we designed, fabricated, and tested a full-custom reservoir computing circuit. This design, the cellular memristive-output reservoir (CMOR), is implemented in 65-nm CMOS technology with integrated front-end-of-the-line (FEOL) resistive random-access memory (ReRAM) used to construct a trainable output layer. We detail the design of this system and present electrical test results verifying its operation and capability to carry out non-linear classifications.

## CCS CONCEPTS

Hardware~Integrated circuits   • Hardware~Application-specific VLSI designs   • Hardware~Emerging architectures   • Hardware~Semiconductor memory

## KEYWORDS

Reservoir computing, cellular automata, in-memory computation, ReRAM, resistive memory

## 1   Introduction

Reservoir computing is a recent approach to machine learning, which attempts to sidestep some challenges of machine learning, such as the complex nature of training time-recurrent systems. To accomplish this, a 'reservoir' is constructed from many interdependent computation units, such as neurons which are randomly connected to one another. Inputs applied into the reservoir then cause it to change its internal behavior. An output layer which examines the state of the reservoir is then the sole feature which is trained to recognize outputs from the reservoir which are related to the task at hand (such as recognizing an image class). The weights and action of the reservoir itself are not explicitly trained.

Several methods of creating reservoirs have been demonstrated, including a literal reservoir of water in one case [1]. The basic requirement for a reservoir is that applied inputs can create a wide variety of complex and evolving states within the reservoir. Additionally, to apply this computing method to temporal tasks, the reservoir must have an 'echo' property in which effects from previous inputs persist in the reservoir before decaying away. To create such systems, neural networks, coupled oscillators, and the focus of this work, cellular automata, have been utilized.

Cellular automata (CA) are interesting mathematical constructions which have potential for reservoir computing. A classic example of a CA is Conway's "Game of Life," in which each cell in a grid follows a set of rules for each element to be declared 'living' or 'dead' in an iterative process. CAs have been demonstrated in software and hardware as reservoir elements, with several different schemes available for introducing echo-states into the computation [6].

Advantages of using CA as a reservoir system include the simplicity of the active element, and the mathematical tools which can be applied to analyze and improve these systems. Additionally, power consumption is low, as a CMOS implementation will only use active power after inputs are changed and the output layer is being read.

ReRAM describes a broad class of memories which have the common characteristic of being two-terminal devices which have internal dynamics that allow their resistance to be modulated or read through specific electrical treatments [4]. One potential application of these devices is to create systems which carry out in-memory calculations.

Using a custom process based on IBM 65nm 10LPe process technology which is fabricated at the SUNY Polytechnic 300mm fab, we have designed and fabricated a CMOS implementation of a CA reservoir integrated with a trainable, ReRAM support-vector machine (RSVM). We overview the architecture of this 'cellular memristive-output reservoir' (CMOR) system, and present initial electrical results from test wafers which confirm its logical operation, programmability, and capability to carry out non-linear classifications.

## 1.1 Elementary Cellular Automata

The fundamental calculation which the cell of an automaton must carry out is that given its internal state and the state of the neighbors to which it is considered 'connected,' it must apply a fixed rule to decide from this information what its own state will be during the next iteration. The repeated application of this operation across all cells in an automaton allows for surprisingly complex dynamics to arise.

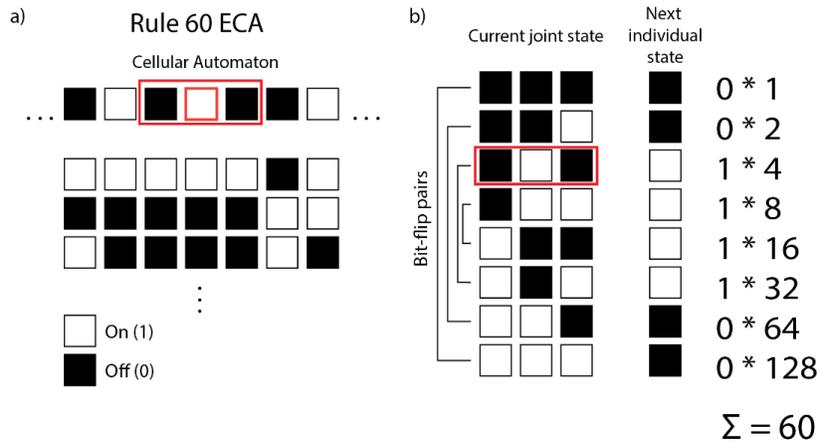

**Figure 1: Illustration of a rule 60 ECA. a) The operation of a 7-cell ECA with periodic boundaries operating under rule 60 is shown over 4 generations. From its initial state (top), following iterations of the automaton change state based on the rule. b) The rule which computes each cell's next state is**

**shown, illustrating how each cell and its neighbors are used to decide that cell's next state. Bit-flipped states lead to an identical next state under this rule, meaning that two inputs which are the bitwise inverse of one another yield an identical next state.**

Elementary cellular automata (ECA) are a class of CA which follow certain restrictions: they exist in one dimension, have binary states, and each cell's state only depends on itself and its 2 nearest neighbors. From this, it is given that each cell has 3 binary inputs: its left neighbor, itself, and its right neighbor (Figure 1a). This yields a total of $2^3$ (8) possible states a cell in an ECA must recognize (Figure 1b). The rule governing the CA must provide a binary response to each of these 8 states, leading to a set of $2^8$ (256) possible rules to govern an ECA. Each rule is commonly summarized by the base-10 (decimal) representation of its responses to each of the 8 input states.

From these well-known rules, subsets have been established within classes that display similar characteristics. Some converge to constant states (class 1), others to repeating patterns (class 2). Others display chaotic characteristics (class 3), and some edge-of-chaos behaviors (class 4) [8]. Each class can provide utility within the context of CA-based reservoir computing, and state-of-the-art systems often combine mixed-rule reservoirs to enhance information processing [5–7].

## 1.2 Support-Vector Machines

A trainable output layer is required in reservoir computing to recognize features within the complex dynamics of the reservoir which are salient to the task at hand. Often, this is implemented by a well-known tool in machine learning, the support-vector machine (SVM). The SVM is a linear classifier which establishes hyperplanes ($\vec{w}$) that attempt to separate points ($\vec{x}$) belonging to one of two classes ($y \in [-1,1]$) [2,3].

$$y_i(\vec{w} \cdot \vec{x_i} - b) \geq 1 \qquad (1)$$

Mathematically, an SVM calculates the dot product of a point ($\vec{x_i}$) with a hyperplane and compares it against a threshold ($b$) to determine its class. Given that in this case, the SVM is operating on the output of a CA, it is effectively using the CA as a 'kernel' to allow it to produce non-linear classifications. This allows it to carry out complex tasks while utilizing a relatively simple architecture.

# 2 Circuit Architecture

## 2.1 Cellular Automata

The cellular nature of a CA-based reservoir system allows for its implementation to be highly compartmentalized and scalable. The elementary unit is a single cell, which then must be connected to its neighbors to form a row. These rows can then be stacked to allow for multiple generations of the automata's state to be computed near-simultaneously. The limiting speed factor in this computation is the propagation of signals within the circuit.

As previously mentioned, the computation each cell must carry out is to decide its next state given its own current state, and the state of its two nearest neighbors. As this requires 3 inputs to select one of 8 outcomes, a 3-bit multiplexer can carry out the computation required by a cell operating under any rule. The multiplexer's selection bits cause it to select one of eight outputs, which can be hard-programmed via the mask design, or if a memory element such as a shift register is included in the cell, dynamically programmed at run-time. This implements the individual cell which form the automaton (Figure 2a).

To build a multi-element automaton, the individual cells are wired into rings of elements, avoiding the need to program specific boundary conditions. The input applied to each cell, representing its current state, is also routed to both of its neighbors. For *n* cells wired to form a ring, a matching *n* bits of input are required (Figure 2b). The entire row's output is applied to another ring representing the next generation. By stacking *m* rings, the state of the ECA over *m* generations is computed and stored (Figure 2c).

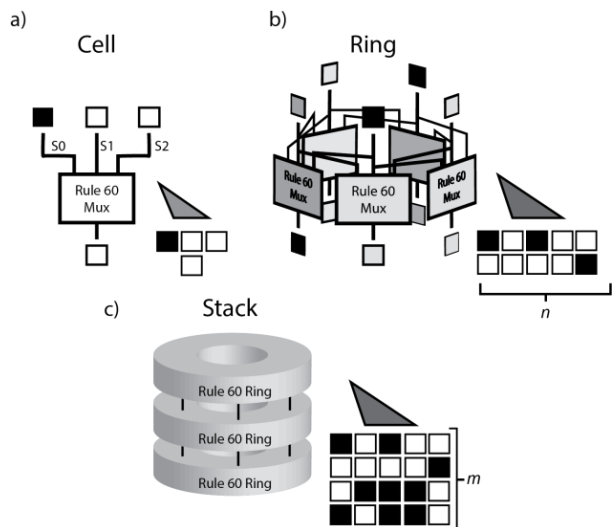

**Figure 2: Illustration of how a multiplexer (mux) implemented cell (a) is arranged into rows (b) and stacks (c) to compute the state of an ECA over multiple generations.**

## 2.2 ReRAM Support-Vector Machine

To carry out a circuit-based implementation of an SVM which can classify the outputs produced by the reservoir, a matrix of 1 transistor 1 ReRAM (1T1R) structures with dimensions ($n \times m$) equal to that of the reservoir is constructed. The binary output of each cell in the automata is wired to the gate of the corresponding 1T1R structure, enabling or disabling current to flow through the ReRAM. The sources and drains of all 1T1R structures are connected to common terminals, wiring them in parallel.

Thus, the SVM's action of calculating the dot product of a point (here, the CA's state) with a hyperplane (the conductivity state of the ReRAM bank) is carried out by using the CA's output to control elements of the ReRAM bank and summing the current flowing through all enabled devices. The measured conductivity is compared against a threshold value to hard-classify the applied input. This implements an analog, ReRAM support vector machine (RSVM).

$$G_\Sigma(x_i) = vec(\mathbf{CA}(x_i)) \cdot vec(\mathbf{G}) \qquad (2)$$

$$sgn(G_\Sigma(x_i) - G_b) = y_{predicted} \qquad (3)$$

Additionally, each individual 1T1R element can be enabled by an external signal, allowing its ReRAM element to be individually addressed and programmed. This allows for the correct conductivity values for the task at hand to be programmed into the memristor bank, storing the classification hyperplane in the collective conductivity of the elements.

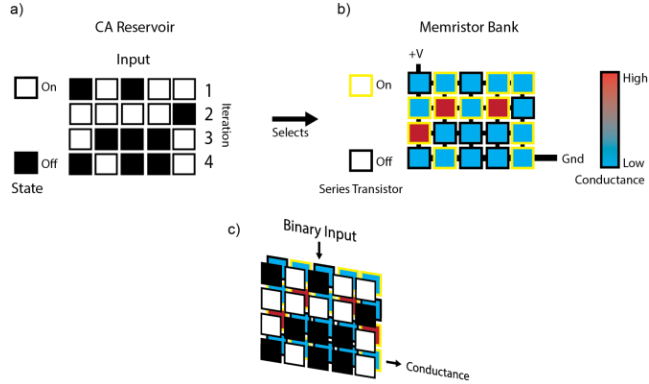

**Figure 3: Illustration of how the integrated support-vector machine operates on the outputs of the reservoir. Signals from cells in the automaton which have a 'high' state (a) enable current to flow through corresponding memristive elements in a memory bank (b). The collective current flowing through the memory bank causes a single conductance value, which varies depending on the applied input (c).**

## 2.3 Design & Fabrication

Prototype revisions of this cellular memristive-output reservoir (CMOR) circuit were included for processing on a multi-project wafer. Designs were laid out and verified using a custom version of the 65 nm IBM 10LPe process design kit (PDK).

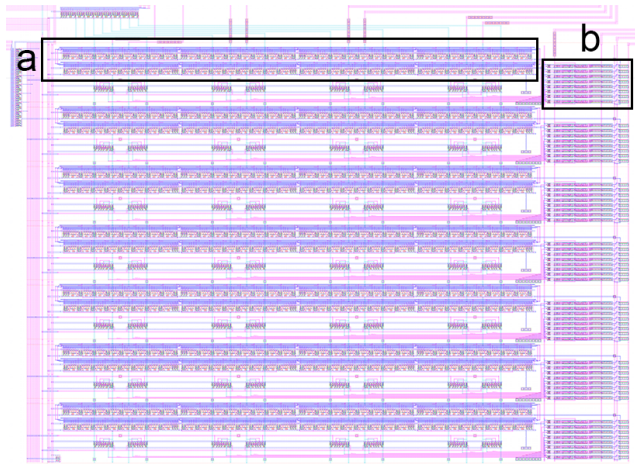

**Figure 4: Layout of a full 8x7 CMOR circuit. Each row of ringed multiplexers (a) forms one generation of the CA, whose state is read by the RSVM (b) to calculate a collective conductance value.**

Changes are made to the standard process to allow the level-1 via (V1), which makes contact between the first and second metal layers (M1 and M2, respectively) to contain a ReRAM element. This is made possible through several changes to the process (Figure 5). To begin with, the metal used for the active-region contact (CA) and first-level metal (M1) is tungsten. Where ReRAM is desired in the V1, titanium nitride (TiN) is then deposited on top of M1 to create studs which serve as the ReRAM bottom electrode (BE). Avoiding the use of copper in these layers enables the use of a front-end of the line (FEOL) atomic-layer deposition (ALD) tool to subsequently deposit a layer of hafnium oxide ($HfO_2$). This oxide serves as the switching layer (SL) of the ReRAM. Next, the SL is covered with a titanium (Ti) film which acts as an oxygen scavenger layer (OSL). As suggested by the name, the OSL reacts with the SL, scavenging oxygen from it and creating a sub-stoichiometric component within the SL. To prevent rapid oxidation of the Ti OSL when exposed to an ambient environment, it is then covered by a titanium nitride (TiN) film which encapsulates the device and serves as the top electrode. Standard copper fills the remaining section of the via, and is used for subsequent metal layers (M2 and up).

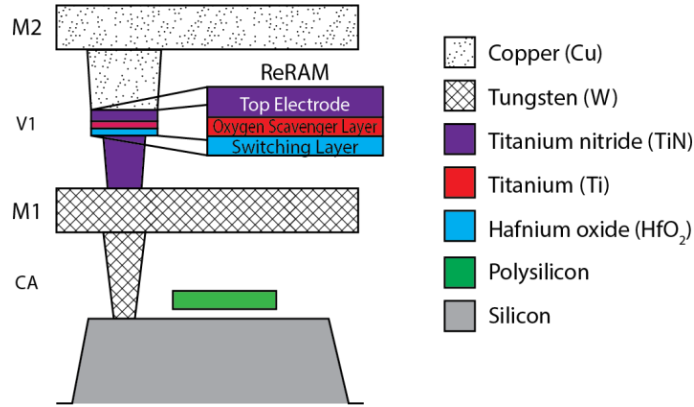

**Figure 5: Illustrated cross-section of a ReRAM device as implemented in our custom process. ReRAM is constructed in level-1 vias (V1) by depositing the requisite layers over a titanium nitride stud.**

All currently-implemented CMOR circuits utilize hard-programmed rules defined at the mask level. As a result, different circuits were constructed to test a variety of ECA rules which have been demonstrated in CA-based reservoir systems. 8 CMOR circuits implementing rules 60, 90, 102, 105, 153, 165, 180, and 195 are currently being fabricated. Each circuit accepts an 8-bit digital input and calculates 7 iterations of the corresponding ECA's state ($n = 8, m = 7$). Each circuit contains a total of 7112 transistors and 56 programmable ReRAM elements. More complex revisions of this design include storage elements which enable dynamic programming of cells and enables larger inputs, but have not yet been fabricated.

## 3  Results

All circuits were fabricated at the SUNY Polytechnic College of Nanoscience & Engineering (CNSE) 300 mm cleanroom facility. Wafer-scale testing was carried out using a probe station equipped with a 12-pin probe card connected via switch matrix (Keithley 707A) to a semiconductor parametric measurement system (Keysight E5270E).

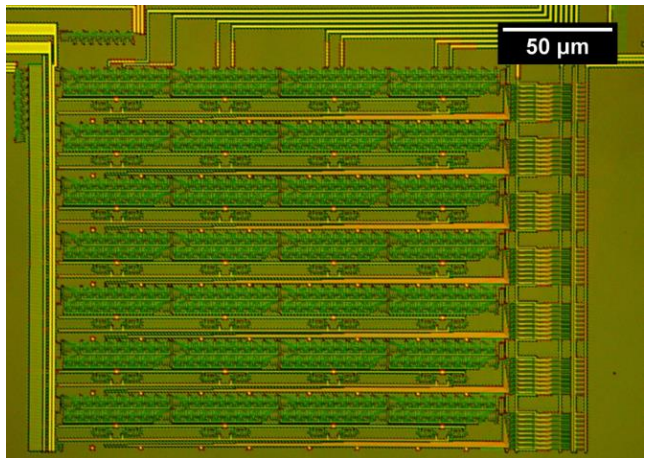

**Figure 6: Light micrograph of an 8x7 CMOR circuit at an intermediate stage of fabrication (M2 level)**

## 3.1 Logic Verification

Each element within the automaton can be selected by its row and column address. This both enables its corresponding weight value in the SVM to be programmed, and routes its current logical state to an output terminal. By applying a digital input and reading the output voltage produced by each cell in the automaton, the logical operation of each implemented rule was checked against those produced by an equivalent software implementation. Each implemented rule displayed the correct behavior (Figure 7).

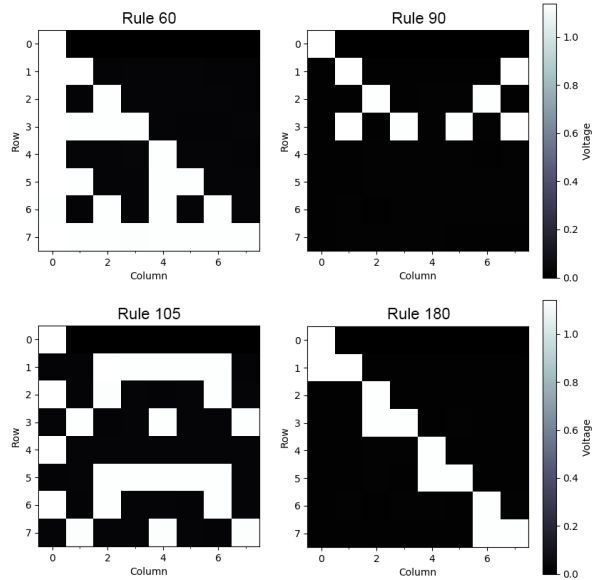

**Figure 7: Electrical test results confirming that each cell in the automata calculates the correct state given a digital input. The automata for each implemented rule was verified as correctly operating.**

## 3.2 Programmability

Having established that the logic underpinning each circuit's operation was correctly operating, we proceeded to test the programmability of the ReRAM support-vector machine (RSVM). Characterization of individual ReRAM elements have previously established that these devices are capable of attaining multi-level resistance states, attaining these values requires a pulse-based programming scheme which was not compatible with the probe-card based interface. Thus, testing focused only on programming ReRAM to a high or low resistance state. This still allows for basic linear and non-linear classifications to be made using the circuit.

### 3.2.1 Single-Element

An individual ReRAM element in a rule 60 CMOR circuit's RSVM was selected by using its row-column address, enabling its control transistor and allowing current to pass through it. This element corresponded to the 2$^{nd}$ iteration's 7$^{th}$ cell (abbreviated (2,7)). A high voltage (3.3 V) was applied to the RSVM's source, and ground to its drain, with a current compliance of 1 mA. The gate voltage for the selection transistors in the RSVM was set at a lower 1.2 V, which limits the size of the filament formed in the ReRAM element. High-voltage (above 1.2V) signals are enabled in the RSVM as it utilizes robust logic gates using a thicker gate oxide than the pure-logic CA section. A current was observed flowing through this single device as it was formed into the low-resistance state (LRS).

By sweeping through all 256 possible 8-bit binary inputs, the dynamic range of the RSVM's conductance after programming the single device was measured. Two major levels of conductivity were observed over all inputs, with the largest changes in conductivity corresponding exactly to the single device in LRS being enabled/disabled. Minor variations in conductivity are created by the other, unprogrammed elements contributing smaller levels of conductivity as they are selected (Figure 8). This is confirmed by examining the mirror symmetry of the measured conductivities, which reflects the phenomenon that bit-flipped states of a rule 60 CA proceed to an identical next state (**Figure 1**b).

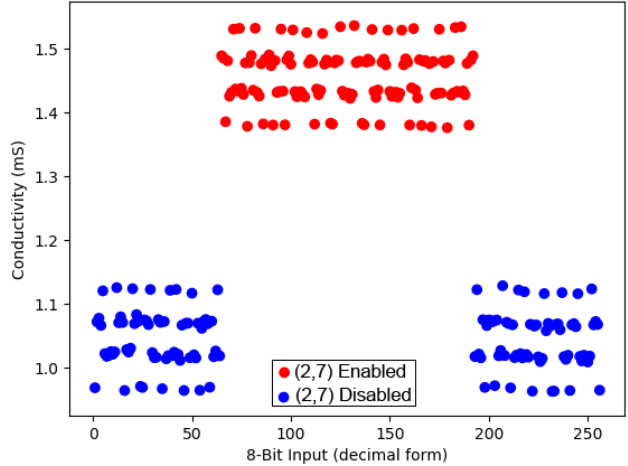

**Figure 8: Conductivity of a rule-60 CMOR circuit's RSVM across all possible 8-bit inputs. The largest shifts in conductivity correspond with the programmed element's control transistor being enabled and disabled.**

Additionally, by programming this element, the circuit exhibits its ability to carry out a non-linear classification. A sub-selection of the input to its 2 most-significant bits ($2^7$ and $2^8$) leads to a non-linear change in conductivity as these inputs iterate through their possible states. Comparing the conductivity of the RSVM over these inputs to a boundary ($G_b$) of 1.2 mS carries out the XOR operation (Figure 9).

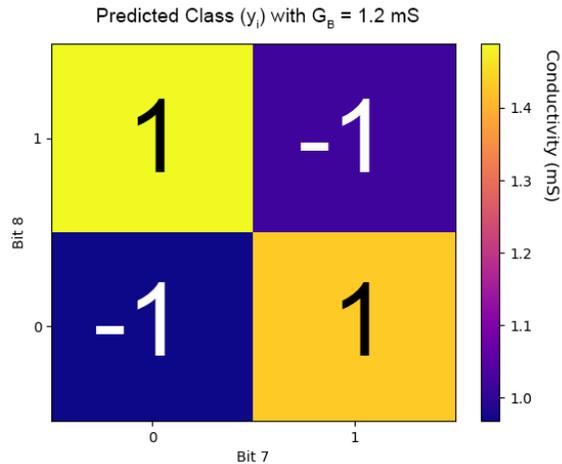

**Figure 9: Example of a non-linear operation (XOR) being carried out by the rule-60 CMOR circuit with 1 programmed element.**

### 3.2.2 Multi-Element

Another element in the 4th iteration's 7th element (element (4,7)) was selected and formed into the LRS using the same electrical parameters. This allowed a different element to contribute significant amounts of conductivity to modulate the collective state of the RSVM. As a result, 3 major levels of conductivity were subsequently observed as each of the 2 elements were enabled and disabled (Figure 10). The slight variation in LRS conductivity between the two programmed elements can be observed in a small but significant difference between the means of the mid-level resistance states where either element (7,4) or (7,2) is enabled (equal variance t-test, α = 0.1%, Figure 11).

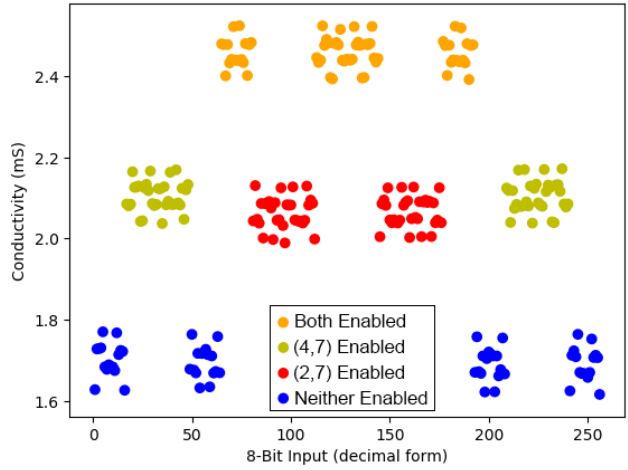

**Figure 10: Conductivity of a rule-60 CMOR circuit's RSVM across all possible 8-bit inputs, after 2 elements in the RSVM have been set to a high conductivity. 3 major levels of conductivity are observed as these elements are enabled and disabled.**

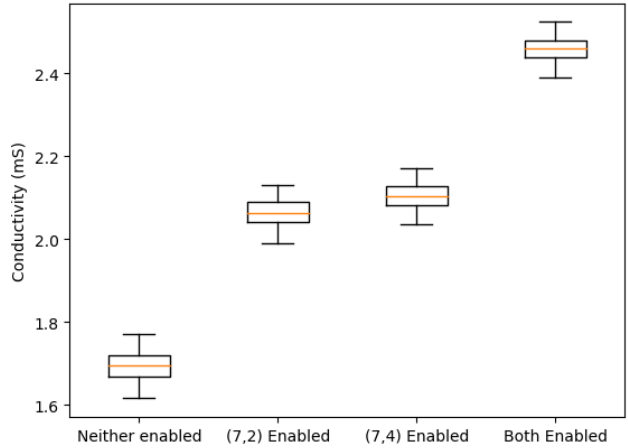

**Figure 11: Distributions of conductivity values sorted by which programmed element was enabled. The differences in conductivity of the (2,7) element and (4,7) element are slight but large enough to make these distributions distinguishable.**

Through these measurements, we verify that the RSVM is operating according to its design. Contributions in conductivity from disabled elements was higher than expected, but as the underlying fabrication process continues to be driven-in, we expect improved performance from the high-voltage control circuitry which is responsible for the RSVM's conductivity modulation.

Having examined both the logic of the CA reservoir and the operation of the RSVM, we have electrically verified the operation of the full CMOR circuit. Initial evaluations of power consumption show that each classification required an average of 0.59 mW. Almost all power was required for either operation of the CA logic (68%) and estimation of the RSVM's conductivity (32%). While the former is a fixed cost of operation, the latter depends both on how the RSVM is programmed and how sensitively its conductivity must be measured. Regularization which penalizes high-conductivity RSVM configurations assist in the former, and the latter likely depends on the degree of separation which can be established between classes which are being distinguished.

## 4  Conclusion

The cellular memristive-output reservoir (CMOR) is a scalable circuit architecture for reservoir computing. In this architecture, the output of a cellular automata (CA) reservoir is used to enable or disable resistive random-access memory (ReRAM) elements which are wired to contribute their conductance in parallel. This creates an integrated reservoir computing circuit with a programmable resistive support-vector machine (RSVM) serving as the output layer.

We have fabricated 8 implementations of this architecture, with each operating under a different CA rule. Comparing simulations and electrical test results, we show that the fabricated circuits reservoirs are correctly operating under each rule. Using a rule-60 based circuit, we program a single ReRAM element in the RSVM and confirm that it is being enabled and disabled by the reservoir. Additionally, having programmed this single element, the RSVM is able to carry out a non-linear classification over its inputs (XOR). We program a second element in the RSVM bank, and confirm that it also contributes its conductivity at the correct time. Through these results, we show that our implementation of CMOR correctly operates in both the reservoir and read-out layer, and can be utilized for non-linear classification.

In future work, we expect the ability to program elements to specific levels via pulsed signals through a modified test setup will increase the system's capability. Additionally, we plan to make the circuit fully scalable to arbitrary sizes through the inclusion of digital memory elements which can be programmed via a standard clocked signal.

## ACKNOWLEDGMENTS
This work was supported by support from AFRL grant FA8750-16-1-0063